# Structure property relationship in (TiZrNbCu)$_{1-x}$Ni$_x$ metallic glasses

Emil Babić[1], Damir Pajić[1], Krešo Zadro[1], Katica Biljaković[2], Vesna Mikšić Trontl[2], Petar Pervan[2], Damir Starešinić[2], Ignacio A. Figueroa[3], Ahmed Kuršumović[4], Štefan Michalik[5], Andrea Lachová[6], György Remenyi[7] and Ramir Ristić[8]

[1]Department of Physics, Faculty of Science, Bijenička cesta 32, HR-10002, Zagreb, Croatia

[2]Institute of Physics, Bijenička cesta 46, P. O. Box 304, HR-10001 Zagreb, Croatia

[3]Institute for Materials Research-UNAM, Ciudad Universitaria Coyoacan, C.P. 04510, Mexico D.F.,Mexico

[4]Department of Materials Science & Metallurgy, University of Cambridge, 27 Charles Babbage Road, Cambridge CB3 0FS, UK

[5]Diamond Light Source Ltd., Harwell Science and Innovation Campus, Didcot, Oxfordshire OX11 0DE, United Kingdom

[6]Institute of Physics, Faculty of Science, P.J. Šafárik University in Košice, Park Angelinum 9, 041 54 Košice, Slovak Republic

[7]Institut Neel, Universite Grenoble Alpes, F-38042 Grenoble, France

[8]Department of Physics, University of Osijek, Trg Ljudevita Gaja 6, HR-3100 Osijek, Croatia

Abstract

The atomic structure, electronic structure and physical properties of $(TiZrNbCu)_{1-x}Ni_x$ ($x \leq 0.5$) metallic glasses (MG) were studied in both the high-entropy ($0<x<0.35$) and the higher Ni concentration range ($x \geq 0.35$). Atomic structure studies performed with X-ray diffraction and synchrotron powder diffraction provided average atomic volumes, structure factors, radial distribution functions, coordination numbers and packing densities. Electronic structure studies performed using photoemission spectroscopy and low-temperature specific heat provided information about the electronic density of states within the valence band and at the Fermi level and also about interatomic bonding and atomic vibrations (from the Debye temperature and the boson peak). Variations of both atomic structure and electronic structure with x showed a clear change for $x \geq 0.35$, which corresponds to a valence electron number $\geq 7.4$. All physical properties, namely thermal stability parameters, Debye temperatures, boson peaks, magnetic, elastic and electronic transport properties change their concentration-dependence for $x \geq 0.35$. The results are compared with those for binary and ternary MGs of the same elements.

## I. INTRODUCTION

"High-entropy alloys" (HEA) are among the latest and probably the biggest challenge in contemporary condensed matter physics and materials science. HEAs are a new type of alloys based on multiple principal alloying components (four or more) in near equi-molar ratios, molar fractions $\leq 0.35$ [1-4]. This design enables research and probable exploitation of a huge number of completely new systems with structures and properties which can hardly be anticipated [5, 6]. Thus, this strategy provides an opportunity to greatly advance our fundamental understanding of the behaviour of complex alloys. Moreover, several important problems in physics, including the localization of electrons and phonons, various percolation phenomena on different crystal lattices and the quantitative distinction between the effects of topological and chemical atomic disorder, can be studied effectively using HEAs. As a result, research on HEAs has led to the preparation of several hundred new alloys, publication of well over a thousand research papers, over ten reviews of the literature (e.g. [6-15]), focus-issues on HEAs in scientific journals and two books [16,17], in just over ten years. The majority of these studies deal with their phase(s), microstructure and mechanical properties, whereas, so far, their physical properties have received relatively little attention, see for instance Ch. 7 in [17], in spite of their potential as functional materials. Especially puzzling is the near absence of experimental studies of the electronic structures of HEAs and of those properties directly related to the electronic structure. In particular, only four studies of their "low-temperature specific heat" (LTSH) have been reported so far [18, 19, 20, 21] and the first "angle-resolved photoemission spectroscopy" (ARPES) studies have been performed by us recently [22]. Theoretical studies of the electronic structure of HEAs, although more abundant than experimental ones, are still insufficient and with a few notable exceptions (e.g. [23, 24, 25,26, 27]) are mainly focussed on the phases of stability and mechanical properties (see for instance Chapters 8-11 in [17]). Almost all HEAs studied so far are metallic alloys in which the itinerant electrons provide a large contribution to the cohesive energy, which renders almost all their properties very sensitive to the electronic structure, e.g. [28] and references therein. So, the scarcity of experimental studies of their electronic structures is quite surprising.

The truly multidisciplinary aspect of HEAs involving researchers from theoretical physics to engineering, has enabled both the breadth and rapid expansion of research on HEAs. Large research efforts in the studies of phases, microstructures and mechanical properties resulted in a wealth of data confirming the conceptual and technological importance of HEAs [6-17]. In particular, HEAs with ultra-high strength and fracture toughness [29], outstanding mechanical properties at high temperatures, excellent soft magnetic properties, high fatigue, wear, corrosion and irradiation resistance, new biomaterials [30] and diffusion barriers, were recently developed [14,17]. An important feature of HEAs is the simplicity of tuning their properties by adjusting their composition and/or phase content. However, the conceptual understanding of their phases, stability and properties is still insufficient. In particular, several semi-empirical criteria for the formation of different phases: single phase solid solution, intermetallic compounds, a mixture of intermetallic compounds and solid solutions, and an amorphous phase (a-HEA) [6-17, 31] have been used to predict hundreds of solid solution HEAs [17, 32, 33]. However, so far only a few dozen stable solid solution HEAs have been confirmed by experiment (Ch. 11 in [17]). Part of the problem is that the phases of HEA depend not only on composition, but also on preparation and the post-processing conditions. This clearly makes it difficult to judge the success of both semi-empirical and theoretical (e.g. [14] and Ch. 8-12 in [17]) predictions of phases of HEAs and it also strongly affects their measured properties.

Therefore, in spite of enormous progress made in almost all aspects of research and conceptual understanding, there is still ample space for important contributions to experimental research on the relationship between their electronic structure, atomic structure and properties, which has been hardly studied so far, and is crucial for deeper conceptual understanding. Further, the studies of HEAs based on the iron group of 3d-transition metals (with the addition of Al, Nb, Sn and metalloids) are by far the most abundant, followed by more recent studies of alloys based on "refractory metals", RHEA, and a small number of studies of other alloys such as those containing "rare-earth metals", noble and normal metals and light elements [14,17]. There are only a few studies of HEAs based on combinations of late 3d and refractory/early transition metals in spite of the fact that in such systems the transitions from a-HEA to simple solid solution HEA, which are conceptually very important, have been reported (e.g. [18] and Ch. 13 in [17]). Indeed, until recently a-HEAs received relatively little attention [15,18] in spite of the fact that they were the first applications of a new alloy design [1,2] and are of crucial importance for the understanding of some features specific to disorder, such as the boson peak [18]. This was probably due to fact that the prediction of "high entropy bulk metallic glasses" (HE-BMG) is probably even more difficult than for conventional binary and multi-component alloys. Further, the critical thickness of HE-BMGs is generally lower than that of conventional BMGs [15]. More recently the situation has started to change and several studies showing the conceptual and technological relevance of a-HEAs appeared [18,34,35,36].

Our previous [18,36,37] and current work on HEAs is focused on three conceptually-important problems in contemporary research on HEAs:

(1) The relationship between the electronic structure, atomic structure and the intrinsic properties of HEAs.

(2) The nature and influence of the transition from HEA to conventional alloys of the same metals, based on one, or at most two, principal alloying components.

(3) The quantitative disentanglement of the effects of structural/topological and chemical disorder by using the same alloy subjected to different preparation or post-processing treatments.

According to a previous report [37] the (TiZrNbCu)$_{1-x}$Ni$_x$ ($x \leq 0.25$) a-HEA system seemed suitable for the study of both problems (1) and (3). Amorphous alloys are suitable for studying (1) because they have a single homogeneous phase and relatively simple electronic structure. In addition, electronic structure and electronic structure-property relations in binary and ternary amorphous alloys of the early (TE) and late (TL) transition metals were found to be particularly simple (e.g. [28, 38] and refs. therein). Further, it was necessary to check whether the HEA design makes some fundamental change to the electronic structure and electronic structure-property relationships or not. Amorphous alloys, especially of the TE-TL type, are also suitable for the study of problem (2). Namely the transition from HEA to conventional alloys with a small number of components, in the same alloy system and without a change of the amorphous phase over to their broad "glass forming composition range" (GFR). As regards problem (3) it was reported [37] that in these alloys, depending on preparation method for instance the cooling rate from the melt, either a- or crystalline alloys with a body-centred cubic (bcc) phase can be obtained, which makes them suitable for disentangling the contributions of topological and chemical disorder to their electronic structure and physical properties.

Here we present the first experimental study to our knowledge of the atomic structure-electronic structure -property relationship in an a-HEA alloy system: $(TiZrNbCu)_{1-x}Ni_x$ ($x \leq 0.5$) for $x$ in both the HEA and Ni-rich ($x \geq 0.35$) concentration range, thus covering problems (1) and (2). Variations of both atomic structure and electronic structure with $x$ show a pronounced change for $x \geq 0.35$ which is reflected in all properties studied, namely thermal stability parameters, Debye temperatures, boson peaks, magnetic, elastic and electronic transport properties all show changes in their concentration dependence for $x \geq 0.35$. The results are compared with those for binary and ternary MGs of the same elements. We also made numerous attempts to prepare crystalline solid solution HEAs (see Experimental) with $x$=0.125 and 0.15, but without success.

## II. EXPERIMENTAL

The ingots of seven alloys in the $(TiZrNbCu)_{1-x}Ni_x$ system with $x$ = 0, 0.125, 0.15, 0.20, 0.25, 0.35 and 0.5 were prepared from high purity elements ($\geq$ 99.98 %) by arc melting in high purity argon in the presence of a Ti getter. The ingots were flipped and re-melted five times in order to ensure good mixing of components. Ribbons with a thickness of about 20 μm were fabricated by melt spinning molten alloys on the surface of a copper roller rotating at a speed of 25 m/s in a pure He atmosphere [18.36]. Casting with controlled parameters resulted in ribbons with closely similar cross-sections (~2 x 0.02 mm$^2$) and thus for x≥0.125 with amorphous phases having a similar degree of quenched-in disorder. Molten alloys with $x$=0.125 and 0.15 were also suction-cast into a water-cooled conical die with a length 50 mm and base diameter 8 mm in order to determine the cooling-rate dependence of the precipitated phase(s) and their contents. (We hoped to find out the conditions for the reported transition from a-HEA to solid solution-HEA with bcc crystalline structure [37].) The as-cast samples were investigated by: (1) XRD using a Bruker Advance powder diffractometer with a CuK$_α$ source, (2) scanning electron microscopy (SEM) using a JEOL ISM7600F microscope with energy-dispersive spectrometry (EDS) capability, and (3) differential thermal analysis (DTA) and differential scanning calorimetry (DSC) using a Thermal Analysis-DSC-TGA instrument. XRD measurements were also performed on crystallized ribbons with $x$=0.125 and 0.15, in an attempt to verify the reported transition from a-HEA to solid solution-HEA [37]. These ribbons were annealed in a pure argon atmosphere [28] for different times (10-60 min) at several temperatures within the temperature range of the first crystallization maximum [36] in the corresponding DSC traces (773-847K). The atomic structure of as-cast samples was also studied using "Synchrotron X-ray powder diffraction" (SXPD) measurements at the I12-JEEP beamline [39] at the Diamond Light Source Ltd., United Kingdom. A piece of a sample ribbon was illuminated with a monochromatic beam of 0.1545 Å wavelength and 0.5 × 0.5 mm$^2$ size for a total time of 240 s. After every sample measurement, the air scattering signal was measured under the same experimental conditions. X-ray radiation of high energy (80.245 keV) was used to cover high $Q$ values of up to 18 Å$^{-1}$ in reciprocal space, giving information about the atomic pair distribution functions of the as-prepared (TiZrNbCu)$_{1-x}$Ni$_x$ alloys. All diffraction experiments were carried out in transmission mode using a flat-panel Pixium RF4343 detector. Precise energy calibration was achieved by collecting diffraction data from a fine powder of CeO$_2$, obtained from NIST, at various standard-to-detector distances. The whole calibration procedure and integration of the two-dimensional X-ray diffraction patterns were performed using the DAWN software package [40].

Thermal measurements were performed with a ramp rate of 20 K/min up to 1550 K. The DTA equipment is regularly calibrated using strontium carbonate and gold standards. This procedure keeps the uncertainty in temperatures derived from DSC-DTA measurements to within ± 5 K. The valence-band structure of as-cast samples was studied with photoemission spectroscopy performed in an ultra-high vacuum chamber equipped with a Scienta SES100 hemispherical electron analyser. The overall energy resolution in the experiments was 25 meV. An unpolarized photon beam of 21.2 eV was generated by a He-discharge ultraviolet source. The samples were cleaned by several cycles of sputtering with 2 keV Ar$^+$ ions at room temperature in order to remove oxygen and other contaminants from the surface. The base pressure during the experiments was below 10$^{-9}$ mbar. Photoemission measurements were also performed on a sample of the alloy with $x$=0.125 that had been crystallized in-situ by holding it at 810 K for 30 minutes. As-cast ribbons were also used for measurements of the LTSH, magnetic susceptibility and mass density $D$ [18]. LTSH measurements were performed in the temperature range 1.8-300 K using a Physical Property Measurement System (PPMS), Model 6000 from Quantum Design Inc. [18, 41]. The magnetic susceptibility of all as-cast and some crystallized samples was measured with a Quantum Design magnetometer, MPMS5, in a magnetic field $B$ up to 5.5 T and temperature range 5-300 K [18,28,36,38]. Since the magnetic susceptibility of all samples showed a very weak dependence on temperature, as is usual in non-magnetic metallic glasses and compounds of early and late transition metals [18,28,38], in the following analysis we will use their room temperature values. The Young´s modulus, $E$, calculated from the relationship $E = Dv^2$, where $v$ is the velocity of ultrasonic waves along the ribbon, was measured both on as-cast ribbons and the same ribbons relaxed for a short time close to the glass transition temperature of a given alloy [18,36,38].

### III. RESULTS AND DISCUSSION

**A. Thermo-physical parameters and sample characterisation**

Because of the vast number of HEAs that can be designed from about eighty stable elements [10, 14] searching for conceptually and technologically interesting compositions by trial and error is clearly inadequate. Therefore, as mentioned in the Introduction several semi-empirical criteria for the formation of different phases: single phase solid solution, intermetallic compounds and a mixture of intermetallic and solid solution phases, and amorphous phase (a-HEA) have been developed [6-17,31-33]. These criteria are based on thermo-physical parameters such as the configurational entropy $\Delta S_{conf}$, the mixing $\Delta H_{mix}$ or formation enthalpy $\Delta H_f$, the average difference in atomic sizes of the constituents δ (as in the Hume-Rothery and Inoue's rules, see e.g. [13,18]), etc. (see Gao et al [42] for an excellent, concise discussion of all semi-empirical criteria and corresponding parameters).

For example, the oldest criterion [43], a two dimensional $\Delta H_{mix}$ (or $\Delta H_f$) - δ plot, where δ is the atomic size mismatch, shows how HEAs evolve with increasing δ and decreasing $\Delta H_{mix}/\Delta H_f$ from solid solution HEAs situated in the region with $\delta \leq 6.6\%$ and -15 kJ/mol$\leq \Delta H_{mix} \leq$-5 kJ/mol to intermetallic and a-HEAs at higher δ and lower or similar $\Delta H_{mix}$ [13,42]. In order to predict the crystal structure type of solid solution HEAs the "valence electron concentration" (*VEC*) criterion was proposed [44]. According to this criterion a bcc phase forms for *VEC* ≤ 6.87, a mixture of bcc and fcc phase forms for 6.87 ≤ *VEC* <8, and a fcc phase appears for *VEC* ≥ 8. The expressions for all these criteria and the definitions of the corresponding parameters can be found in reviews of the literature [6-17,42] as well as in our previous papers [18,36].

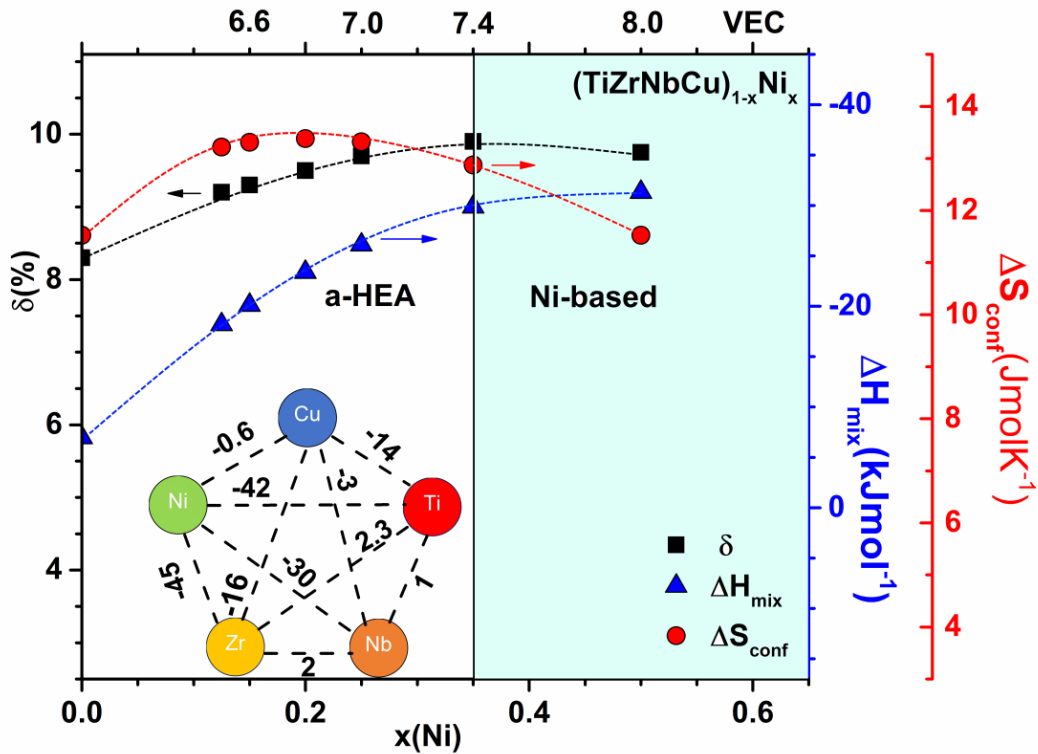

FIG. 1. Thermo-physical parameters of (TiZrNbCu)$_{1-x}$Ni$_x$ alloys vs. x and valence electron count, *VEC* (upper abscissa). Left scale: atomic size mismatch $\delta$, right scales: mixing enthalpy $\Delta H_{mix}$ (first right scale), and configurational entropy $\Delta S_{conf}$ (second scale). The inset: $\Delta H_{mix}$ between constituent elements [32]. Lines are guide for eye.

In Figure 1 we show the variation with concentration x of parameters $\Delta H_{mix}$, $\delta$ and $\Delta S_{conf}$ of $(TiZrNbCu)_{1-x}Ni_x$ alloys. Here, the HEA region of $x$ is distinguished from the Ni-rich one ($x \geq 0.35$) by a different colour. As already noted [18,36] the values of the parameters $\Delta H_f$ ($\Delta H_{mix}$) and $\delta$ of all alloys are well outside the $\Delta H_{mix}$-$\delta$ region in which solid solution HEA form. (More recent and sophisticated criteria such as those in [32,33,42] corroborate this conclusion). Regardless of the reliability of $\Delta H_{mix}$-$\delta$ [13] and other [32,33] criteria, this conclusion is consistent with our difficulties in trying to convert our alloys with low $x$ to solid solution HEAs. Further, the contribution of the ideal $\Delta S_{conf}$ to the free energy is considerably smaller than that of enthalpy [18,36] even at the liquidus temperature $T_l$, so that its effect on phase formation may not be large, as was confirmed by XRD patterns and DSC/DTA results [18,36]. In particular, all melt-spun ribbons with x>0 were fully amorphous, whereas their crystalline counterparts, suction cast rods [37] and bulk conical samples were multiphase.

The inset of Figure 1 shows that the strong interatomic bonding between TE and TL components, especially between Zr and Ti with Ni, are responsible for the small overall values of $\Delta H_{mix}$ (the values of $\Delta H_{mix}$ in the inset are taken from [32]). These strong interactions are likely to give rise to "chemical short range order" (CSRO) in the amorphous phase [38] and intermetallic compounds in the crystalline state [28]. Simultaneously, they seem to stabilize amorphous phases, since the alloy with x=0 did not vitrify [18]. Similarly, the difference in atomic radii [45] between TE and TL metals is responsible for rather large values of $\delta$ which show a maximum at $x$=0.35.

Since *VEC* is proportional to $x$ of our alloys, the upper abscissa enables one to follow the variations of the parameters shown in Figure 1 as a function of *VEC*. We note that *VEC*≤7 for $x$≤0.25 (HEA region) and reaches 8 for $x$=0.5 (Ni-rich region). Thus, provided the correlation between *VEC* and crystal structure of numerous solid solution HEAs [42,44] also applies to TE-TL type of multicomponent amorphous alloys, the transition from the HEA to Ni-rich concentration range in our alloys could coincide with a change in the "atomic short range order" (SRO), from say bcc-like to mixed bcc-fcc-like and finally to fcc-like. Indeed, an approximate analysis of the first maxima in the XRD patterns of amorphous $(TiZrNbCu)_{1-x}Ni_x$ alloys with $x \leq 0.25$ indicated bcc-like local atomic arrangements [18], whereas a more recent study of the XRD patterns of the alloys with $x=0.35$ and 0.5 (Ni-rich concentration region) indicated deviations from the behaviour expected for bcc-like atomic arrangements [36]. In particular, the average lattice parameters $a$ calculated from the XRD patterns [38] by assuming bcc-like local atomic arrangements decreased linearly with $x$ for $x \leq 0.25$ (as expected from Vegard`s law) but $a$ showed a strong positive deviation for $x \geq 0.35$ (Figure 2 in [36]). Before discussing in some detail this unusual finding, we will briefly overview the main results of thermal studies and compositional/homogeneity characterisation of our samples (given in more detail in [18, 36]).

The detailed description of SEM/EDS studies performed on our as-cast samples in the HEA composition region, which included SEM images and EDS mapping of the distributions of constituent elements was given in [18]. Briefly, it was found that the EDS compositions were the same as the nominal ones to within ±1-2 at % and that the distributions of all constituent elements were random down to the sub-micrometer scale. We note that suction cast rods of $(TiZrNbCu)_{1-x}Ni_x$ alloys [37], having nominally the same compositions as our a-HEAs, solidified to form dendritic microstructure and EDS analysis showed that these dendrites, of size 2-5 µm, consisted mainly of Nb. Indeed, quite often the distribution of constituent elements in HEAs is uneven, and this sometimes occurs even in alloys showing solid solution HEA behaviour in their XRD patterns [7]. We performed the same type of SEM/EDS study on our samples with higher Ni content ($x \geq 0.35$). These samples also had a random distribution of constituent elements and their EDS compositions agreed well with nominal ones [36].

The detailed DSC/DTA study of all samples [36] confirmed the XRD results that as-cast ribbons with $x \geq 0.125$ were amorphous and provided their glass transition ($T_g$), crystallization ($T_x$), melting ($T_m$) and liquidus ($T_l$) temperatures. All alloys showed quite complex crystallization behaviour with two distinct exothermic maxima for $x \leq 0.25$ [36,37] and three maxima in the Ni-rich alloys [36]. For the sake of simplicity, we analysed only the first crystallization event (allegedly associated with the precipitation of a bcc-HEA phase for $x \leq 0.2$ [37]) which determines the stability of the amorphous phase and the width of the super-cooled liquid region, $\Delta T_x = T_{x1} - T_g$. Thermal stability parameters $T_{x1}$ and $T_l$ showed qualitatively the same variations with $x$, a rapid increase for $x \leq 0.25$ and a tendency to saturate in the Ni-rich region (Figure 4 in [36]). We note that the observed change in concentration dependence of $T_x$ and $T_l$ coincides with that in the average lattice parameters of a bcc-like atomic arrangement (Figure 2 in [36]). Since $T_x$ and $T_l$ are associated with the strength of interatomic bonding, it seems that a change in SRO at elevated Ni-contents affects the interatomic bonding too. The enthalpies of crystallization $\Delta H_{c1}$ and the widths of supercooled liquid region showed similar variations with $x$ as those of $T_{x1}$ and $T_l$. In particular, $\Delta T_x$ increased from about 40 K at low $x$ to about 100 K on the more Ni-rich side. The reduced glass transition temperature $T_{rg}$ (which is frequently employed as a criterion for glass forming ability, e.g. [28]) showed a modest magnitude around 0.52 (thus indicating modest GFA) and rather little variation with $x$ (Figure 2 in [36]). Since we experienced some difficulties in preparing the fully amorphous alloys with the lowest and the highest Ni-contents and were unable to vitrify the alloy with x=0, the GFA criteria $T_{rg}$ and $\Delta T_x$ [28] may not be applicable to present alloys [36].

**B. Atomic structure**

As already noted in the previous section, the XRD patterns of amorphous alloys, in addition to showing the amorphous nature of a particular alloy, can also provide some insight into the local atomic arrangements, the average atomic volumes and the average atomic packing fractions (e.g. [38,46]). The corresponding procedures were previously [47,48] used by us in order to correct the $N(E_F)$ of a hypothetical fcc-phase of pure Zr [49] , by providing a better estimate for the corresponding atomic volume, and were also used in order to determine the atomic volume of amorphous copper and the average atomic packing fractions of amorphous Cu-Hf alloys [38]. We note that in all these cases the atomic volumes determined by using XRD patterns agreed quite well with those obtained from the experimental mass-density [50].

In particular, from the modulus of the scattering vector $k_p$, corresponding to the first maximum in the XRD pattern [18,36], $k_p=4\pi sin\theta/\lambda$ ($\theta$ is the Bragg angle and $\lambda$ is the wavelength of the X-ray radiation) one can calculate the average nearest neighbour distance [46]:

$$d = \frac{7.73}{k_p} \quad . \tag{1}$$

By assuming an approximate crystal structure of the local atomic arrangement, one can calculate the corresponding average lattice parameter $a$ and the average atomic volume, $V$. In particular, for a bcc-like local atomic structure $a_{bcc}=2d/3^{0.5}$ and $V_{bcc}=a^3/2$, whereas for the fcc-like atomic SRO $a_{fcc}=2^{0.5}d$ and $V_{fcc}=a^3/4$. The variation of $a_{bcc}$ of all our alloys with Ni-content was shown in Figure 2 of [36]. Whereas $a_{bcc}$ for alloys with $x \leq 0.25$ ($VEC \leq 7$) follows Vegard's law for a bcc crystal structure quite well (making allowance for somewhat lower mass-density of amorphous alloys [36]), the $a_{bcc}$ data for alloys with more Ni ($VEC>7$) shows a strong upward deviation from the Vegard's law.

Here, in Figure 2 we show the corresponding $V_{bcc}$ values for all our alloys as a function of Ni-content $x$ and $VEC$ (upper abscissa). The variation of $V_{bcc}$ is qualitatively the same as that of $a_{bcc}$ [36], $V_{bcc}$ tends to "saturate" for $x \geq 0.35$ at values which are considerably larger than those predicted by Vegard's law. The values of $V_{fcc}$, obtained from XRD patterns by assuming an fcc-like local atomic structure, for $x \geq 0.35$ and also shown in Figure 2 seem to agree better with the expected, approximately linear, variation of atomic volumes with concentration. This is a rather general feature of metal-metal type amorphous alloys [38, 50, 51]. Therefore, the variations of both average lattice parameters [36] and atomic volumes with $x$ (Figure 2) indicate a change in atomic SRO in alloys with more Ni ($x \geq 0.35$, $VEC \geq 7.4$).

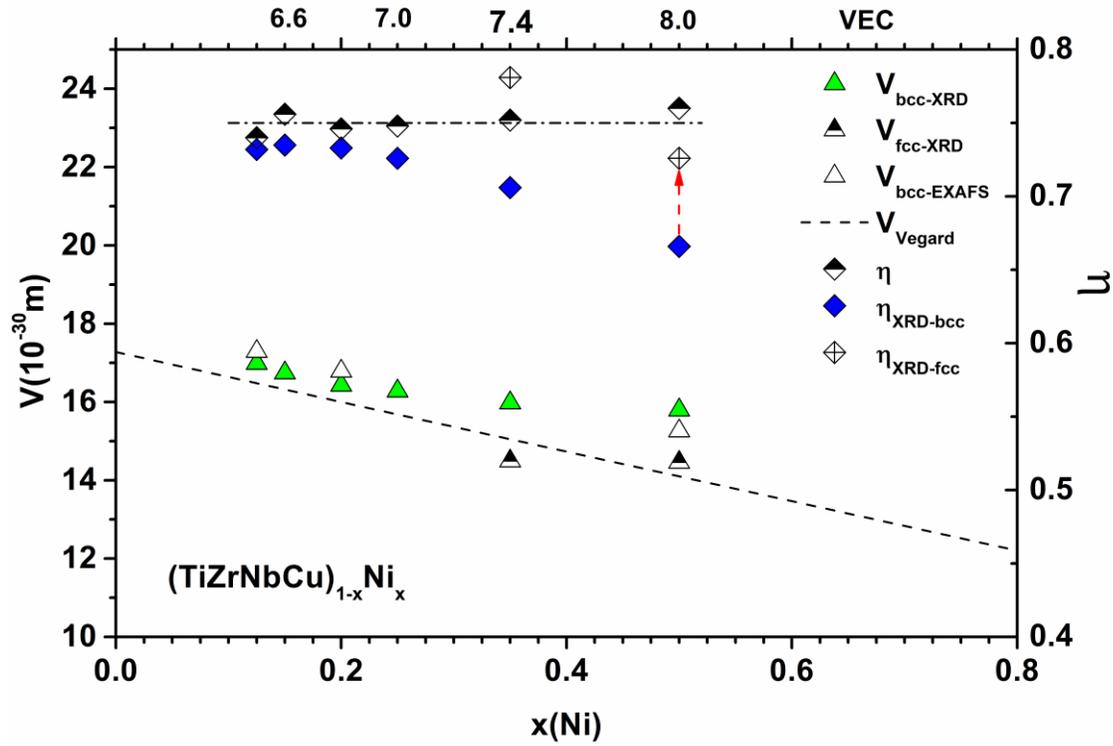

FIG. 2. Atomic volumes and atomic packing fractions of (TiZrNbCu)$_{1-x}$Ni$_x$ alloys vs. $x$ and *VEC* (upper abscissa). Left scale: Atomic volume *V*; right scale: packing fractions $\eta$. Lines are guide for eye.

From the atomic (or molar) volumes one can also calculate the average local atomic packing fractions, APF [51]. In particular, the APF for an amorphous alloy, $\eta_a$, can be calculated from the expression [51]:

$$\eta_a = \frac{\sum_k \eta_k^0 x_k V_k^0}{V_a} , \qquad (2)$$

where $\eta_k^0$ is the APF of the k-th constituent of the alloy in its crystalline state, $x_k$ and $V_k^0$ are its molar fraction and molar volume in the crystalline state respectively, and $V_a = M/D$, where *M* is the molar mass and *D* the mass density that was calculated from the atomic volumes of amorphous Ti, Zr, Nb, Cu and Ni in [50]. As seen in the upper part of Figure 2 these values of $\eta_a$ are approximately constant, as could be expected because $V_a$ obeys Vegard's law, and fairly high, $\eta_a \approx 0.75 \pm 0.01$. The APF obtained by using $V_{bcc}$ values, (thus $V_{bcc}$ from Figure 2 instead of $V_a$ ) $\eta_{bcc}$, is also shown in the upper part of the same Figure. $\eta_{bcc}$ is like $\eta_a$ approximately constant but is a little lower (lower D [18]) than $\eta_a$ for $x<0.25$, but rapidly decreases to $\eta_{bcc}=0.67$ for $x=0.5$. Thus, the variations of both $V_{bcc}$ and $\eta_{bcc}$ indicate changes in atomic SRO for $x\geq0.35$.

However, by replacing $V_{bcc}$ with $V_{fcc}$ for x=0.5 one recovers the expected approximately constant APF, specific to amorphous alloys of the metal-metal type [38,51]. Thus, the variations of all parameters derived from XRD patterns (Figure 2 and [36]) indicate a progressive change in local atomic arrangements accompanying the transition from HEA to the concentration region with higher Ni content. This change apparently affects the thermal parameters (section A and [36]) and probably the interatomic bonding in the alloys studied.

However, Guinier's procedure [46] for the determination of the average distance between the nearest neighbours in amorphous alloys from XRD patterns has been vigorously criticized (e. g., [52]). Therefore, we recently started SXPD measurements [53] which provide a more direct insight into the possible changes in the atomic SRO, obtained from the structure factors $S(Q)$ and radial distribution functions $RDF(r)$ as well as from a more reliable determination of $d$ . The total X-ray $S(Q)$ was obtained from integrated raw intensity data, $I^{raw}(Q)$, using the procedure described elsewhere [54,55]. Briefly, $I^{raw}(Q)$ was corrected for background (air scattering), self-absorption, fluorescence and Compton scattering and then scaled and normalized into electron units using the high-angle region method [56]. The corrected intensity, $I^{cor}(Q)$, was used to calculate $S(Q)$ by applying the Faber-Ziman formalism [57]. All the corrections mentioned above were obtained using the PDFGetX2 program [58].

The $S(Q)$ curves (not shown) consisted of a strong initial peak, followed by a series of broad, damped oscillations extending up to $Q=16$ Å$^{-1}$, which together rule out any crystallinity in our samples with $x\geq0.125$. The position of the first peak shifted rapidly with $x$ to higher $Q$ for $x\leq0.25$, but the shift slowed down for $x=0.5$. A more drastic change occurred in the second, split maximum of $S(Q)$ which is probably more directly related to the atomic SRO than the first one (e.g. [59]). In particular, its asymmetry, the difference in the magnitudes of its first and second part (higher $Q$), was strongly reduced at $x=0.5$ with respect to those observed for $x \leq 0.25$.

The radial distribution functions $RDF(r)$ were obtained through sine Fourier transformation of $S(Q)$:

$$RDF(r) = 4\pi r^2 \rho(r) = 4\pi r^2 \rho_0 + r \frac{2}{\pi} \int_0^\infty Q[S(Q) - 1] \sin(rQ)\, dQ \qquad (3)$$

where $\rho(r)$ and $\rho_0$ are the local and average atomic number densities, respectively, and $r$ is the radial distance.

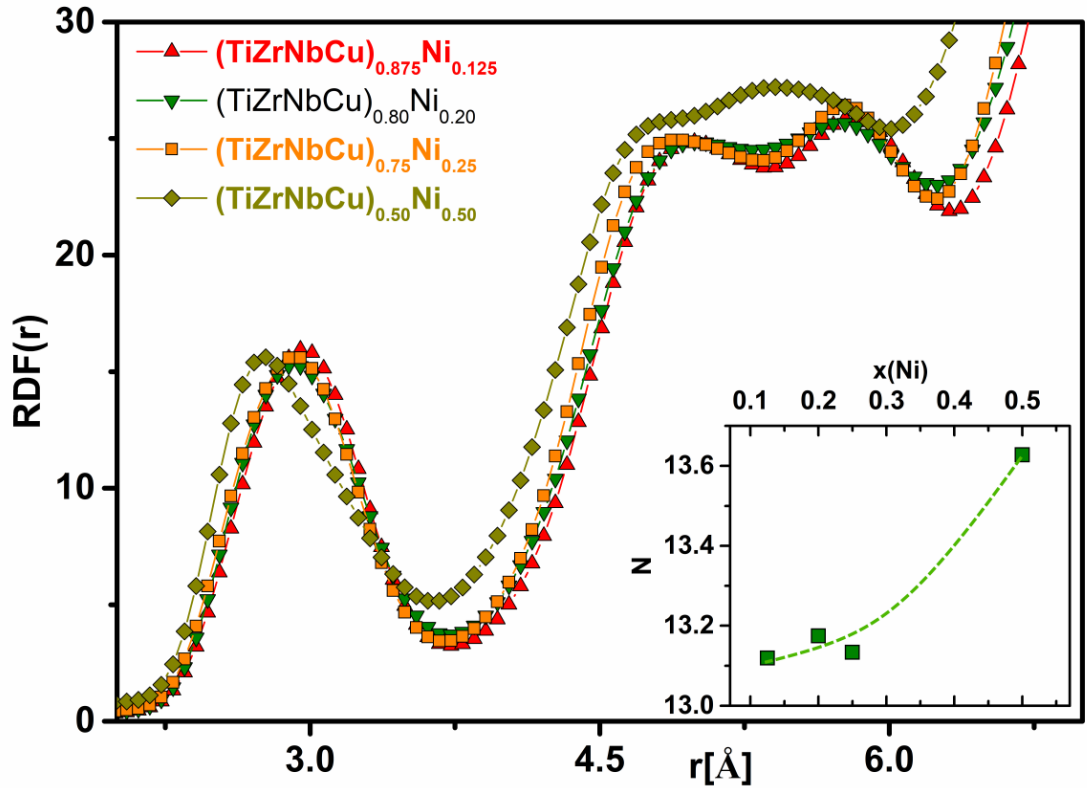

FIG. 3. Radial distribution functions $RDF(r)$ vs. $r$ for (TiZrNbCu)$_{1-x}$Ni$_x$ alloys [53]. The inset: corresponding coordination numbers $N$ vs. $x$. Dashed line in the inset is guide for eye.

$RDF(r)$ curves shown in Figure 3 confirm a strong change in both atomic short- and medium-range order (MRO) occurring for $x=0.5$. In particular, the shape of the first maximum (SRO) becomes strongly asymmetric on the right side of the peak, while in the second, split maximum, the magnitude of its second part (higher *r*) is strongly enhanced with respect to that of its first part. From the positions of the first peaks we determined corresponding interatomic distances and average bcc-atomic volumes which agree well with those obtained from XRD patterns as demonstrated in Figure 2. From *RDF(r)* we also calculated the average coordination number, *N* (the number of atoms in a spherical shell between radii $r_1$ and $r_2$ around an average atom):

$$N = \int_{r_1}^{r_2} RDF(r)dr. \qquad (4)$$

The variation of *N* for the cut-off $r_2$ at the minimum between the first and second maximum in Figure 3 is shown in the inset to Figure 3. A strong increase of *N* is observed at *x*=0.5, confirming the change in SRO. We note that *N*≥13 is usual in multicomponent metallic glasses [59]. Since for high $r_2$ *N* may have some contribution from the second atomic shell [52], we have checked the variations of *N* for somewhat lower values of $r_2$ and obtained qualitatively the same variations with x. In particular, for a cut-off at 0.95 $r_2$ we obtained *N* increasing from 12.4 for *x*≤0.25 to 12.8 at *x*=0.5. Therefore, the initial results of SXPD provide strong support for a change in atomic arrangements at *x*=0.5 (*VEC*=8). This change is accompanied by an increase in *N*, which is reminiscent of the proposed correlation between *VEC* and crystal structure in HEAs [42,44]. Due to the correlation between the local atomic order and electronic structure we expect strong changes in electronic structure in the region with more Ni as described in the next section.

**C. Electronic structure**

As already noted in the Introduction and in several recent papers (e.g. [18,25,28,36,38,60], for metallic systems, regardless whether they are amorphous or crystalline, the electronic structure controls practically all their intrinsic properties (those that are hardly affected by the exact preparation and/or post-processing conditions). The importance of electronic structure in understanding the properties of alloys probably shows up the best in the case of amorphous TE-TL alloys. Soon after the discovery of these MGs "the photoemission spectroscopy" (PES) revealed the split-band structure of these alloys with TL metals having a full or nearly full d sub-band for which the electronic *DoS* at the Fermi level, $N_0(E_F)$, is dominated by TE d-states [61]. Thus, the effect of alloying with TL is approximately described by a dilution of a-TE [62] which simplifies the explanation of the linear variations of most properties of these MGs with TL content [38,47,62]. Furthermore, $N_0(E_F)$ values of TE-rich alloys (determined from LTSH) were higher than those of stable hexagonal close-packed crystalline phases of corresponding TE [63] and were close to those calculated for hypothetical fcc structures of TEs [49]. High $N_0(E_F)$ in TE-rich MGs leads to enhanced superconductivity and magnetic susceptibility [47], but also to weaker interatomic bonding, thus to lower elastic moduli and thermal stability [64]. The combined studies of PES and ab-initio calculations were also performed on Zr-based multicomponent MGs and showed that $N_0(E_F)$ is dominated by TE d-electrons as in binary MGs (e.g. [65]). We note that a combination of PES, LTSH and ab-initio theory is the best in order to fully comprehend the electronic structure. In particular, ordinary PES and LTSH experiments reveal the variation of the total *DoS* with energy (PES) and give accurate value for $N_0(E_F)$ (LTSH), but they cannot provide accurate contributions of the alloying elements (*pDoS*) to these quantities. Theory can in principle provide all these quantities as well as the probable local atomic structure (including chemical short range order [65]) but its results are often limited by the rather small size of the sample and approximations involved in a given calculation (e.g. [38, 60,65]).

Accordingly, we performed combined PES [22, 66] and LTSH [18,36] studies of a- $(TiZrNbCu)_{1-x}Ni_x$ alloys with *x* covering both HEA and regions with higher Ni content. The results of PES in a-HEA region are illustrated in Figure 4 and compared with those for binary and ternary Zr-based MGs [67]. The recorded "ultraviolet photoemission spectrum" (UPS) reflects the *DoS* within the valence band of amorphous $(TiZrNbCu)_{0.8}Ni_{0.2}$ alloy. The spectral maximum at 3.5 eV below the Fermi level, associated with the Cu-3d states, is well resolved and separated from the low-energy part of the spectrum. From the comparison with the spectrum of amorphous (Ni$_{0.33}$ Zr$_{0.67}$)$_{0.85}$ Cu$_{0.15}$ and Ni$_{0.33}$ Zr$_{0.67}$ alloys [67], shown in the lower part of Figure 4, it is evident that the low-energy part of the spectrum has contributions from Ni-3d and Zr-4d bands [67]. It appears that most of the spectral intensity between 1 eV and 2 eV below $E_F$ comes from the *DoS* associated with Ni-3d bands [68,69]. The Ni contribution to the intensity at the Fermi level is not negligible and is likely to increase at high Ni-content [70]. The Zr-4d band is expected to contribute to the *DoS* mainly at the Fermi level [71] as is the Ti-3d band, as judged from the spectrum from pure Ti [72] and an amorphous $(Ni_{0.33}Zr_{0.67})_{0.85}Ti_{0.15}$ alloy [67]. The Nb-4d band contributes to the spectral intensity at the Fermi level and around 1.2 eV [73] but possibly to a lesser extent than Ni [74]. The s-p bands of all constituents span a larger energy range and generally contribute less to the photoemission intensity. Therefore, we could not extract their contribution to the DoS from the spectra of the alloys. The effects of increasing Ni content x and of crystallization of the sample with $x=0.125$ on photoemission spectra, have also been studied [66]. The present results add to mounting evidence that photoemission spectroscopy is an efficient tool for assessing the contributions of individual constituents to the electronic properties of 3d and 4d transition metal based HEA. Indeed, the insight obtained from PES will be helpful for the interpretation of LTSH, magnetic susceptibility and superconductivity data for the same alloys.

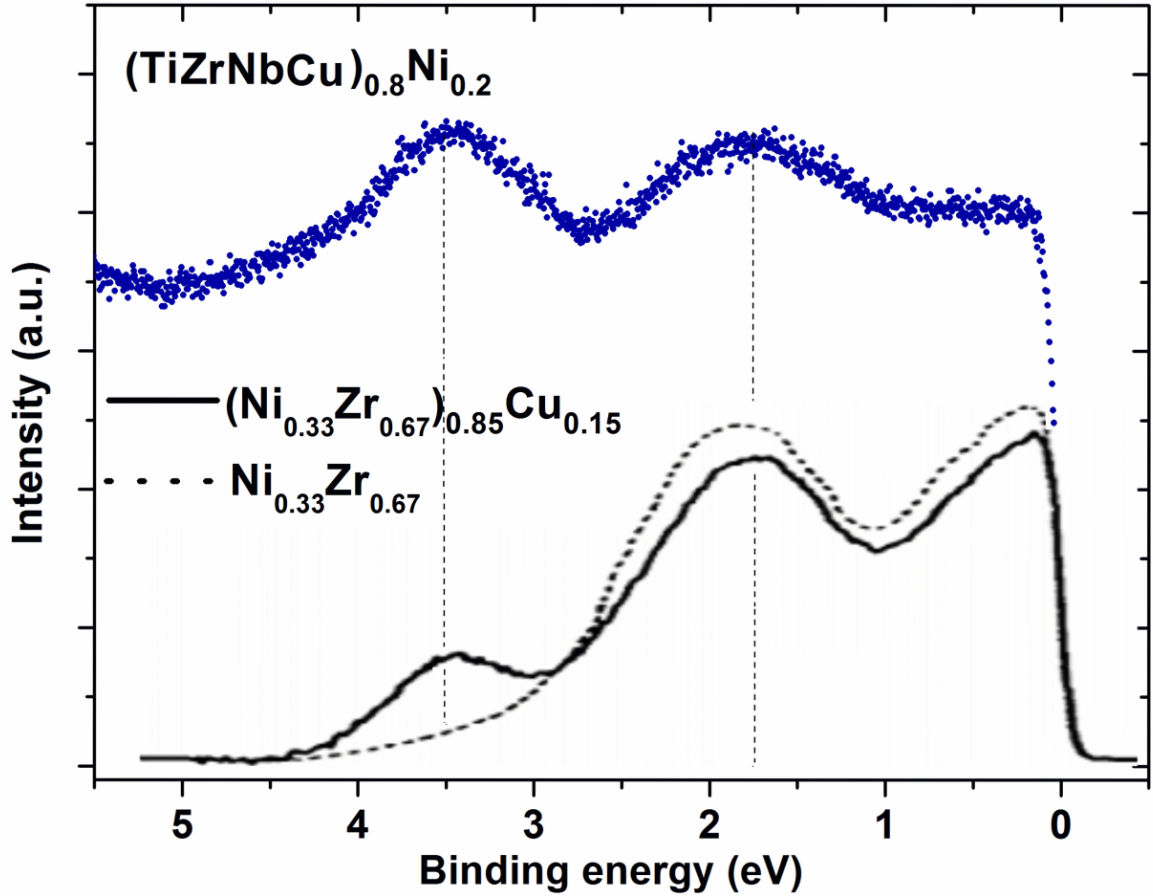

FIG. 4. UPS spectra of (TiZrNbCu)$_{0.8}$ Ni$_{0.2}$ (blue dots, [66]) and for comparison, of (Ni$_{0.33}$ Zr$_{0.67}$)$_{0.85}$ Cu$_{0.15}$ and Ni$_{0.33}$ Zr$_{0.67}$ from ref. [67].

As noted above, LTSH provides a quantitative insight into the *DoS* at $E_F$ which controls the physical properties of metallic systems. In particular, the Sommerfeld coefficient of the linear term in LTSH is given as [18,36,38]:

$$\gamma = \frac{\pi^2 k_B^2 N_0(E_F)(1+\lambda_{e-p})}{3} \text{ ,} \qquad (5)$$

where $k_B$ is the Boltzmann constant and $\lambda_{e-p}$ represents the electron-phonon enhancement of the *DoS* at $E_F$, $N_0(E_F)$. Since the local atomic arrangements determine the ES of alloys (e.g. [28]) we expect a change of $N_0(E_F)$, thus also *γ* of our alloys, in going from a-HEAs ($x \leq 0.25$) to conventional amorphous alloys with higher Ni content ($x \geq 0.35$).

In Figure 5 we compare the variations of *γ* with TL content in our alloys [36] with those in several binary and ternary a-TE-TL alloys composed from the same TEs and TLs [75,76,77,78,79]. In Figure 5 we plot *γ* vs. total TL-content (that of Cu and Ni) in our alloys, since in binary TE-TL amorphous alloys the variations of *γ* in the alloys with TL=Cu and Ni are very similar, thus it is the total TL content which causes a decrease of *γ* in alloys containing both, Cu and Ni. (If we plotted *γ* vs. Ni-content only, as was done in [36], this will only shift the data in Figure 5 towards the left but will not change the overall variation with $x$). As seen in Figure 5, in our alloys $\gamma$ values initially decrease with $x$ as in other a-TE-TL alloys, but they saturate for $x \geq 0.51$, this does not occur in binary alloys at similar TL contents. In particular, $\gamma$ of binary alloys seems to follow a linear decrease with $x$ throughout the $x$-range explored ($x \leq 0.70$). Possible exceptions are a-Zr-Ni alloys, which seem to show slower variation of $\gamma$ with $x$ for $x > 0.60$ ($VEC > 7.5$) [63]. In these alloys this change is associated with a strong chemical short range order (CSRO) effect [38, 50, 63,65] and the change in the position of $E_F$ with respect to the two sub-bands in the *DoS*, as evidenced by Hall effect and thermopower measurements [80]. Ab-initio studies of the atomic and electronic structure of binary amorphous alloys of Ni with Ti, Zr , V or Nb performed by Hausleitner and Hafner [65] corroborated these findings. They found an increase in CSRO and a change in local atomic order at high Ni contents. Simultaneously, d-band of Ni broadened and shifted towards the Fermi energy. These effects could also affect the variation of $\gamma$ with Ni content in our alloys, but preliminary measurements of the Hall effect in a-HEAs do not seem to indicate such band-crossing [81] for $x<0.5$. As regards its magnitude, $\gamma$ of our alloys with $x=0.35$ agrees rather better with those for a-Zr-Cu, Ni alloys than with those for a-Ti-Cu, Ni alloys, in spite of the sizable Ti-content. This may be due to the Nb content in our alloys which reduces both their Zr and Ti contents and, as seen in Figure 5, addition of Nb strongly suppresses $\gamma$ in a-Zr-Nb-Ni alloys [79].

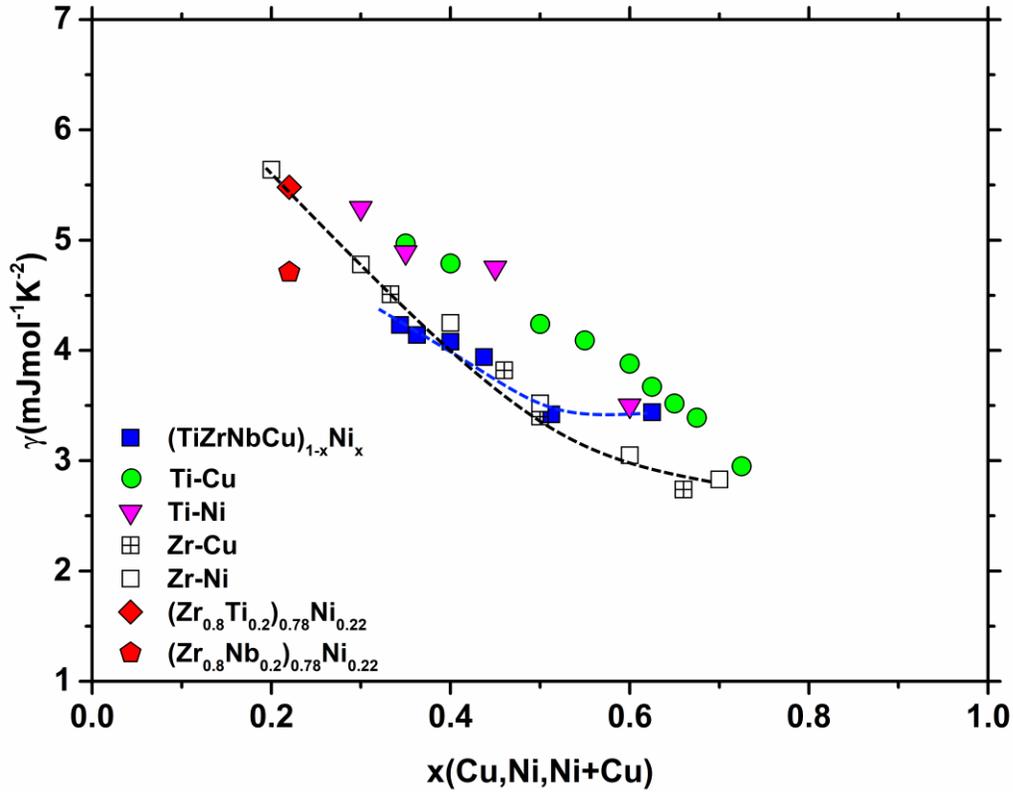

FIG. 5. Variation of electronic coefficient $\gamma$ vs. $x$ for $(TiZrNbCu)_{1-x} Ni_x$ [36] and some binary and ternary amorphous alloys. Data for Ti-Cu are from [75], for Ti-Ni from [76], for Zr-Cu from [77] for Zr-Ni from [78], for $(Zr_{0.8}Ti_{0.2})_{0.78}Ni_{0.22}$ and $(Zr_{0.8}Nb_{0.2})_{0.78}Ni_{0.22}$ from [79]. Dashed lines are guide for eye.

Since $\gamma$ depends on the *DoS* which is enhanced by the electron-phonon interaction it cannot be used in order to prove that the change in $\gamma$ for $x \geq 0.35$ is due to $N_0(E_F)$. However, in superconducting transition metal alloys one can use the McMillan expression [82] in order to disentangle $N_0(E_F)$ from $\lambda_{e-p}$. By using the results of recent measurements of the superconducting transition temperatures, $T_c$ [81], we have calculated the values of $N_0(E_F)$ shown in Figure 6, which confirm a change in $N_0(E_F)$ for $x \geq 0.35$. Thus, our study of the electronic structure is consistent with that of atomic structure in Section B in that a change of electronic structure accompanies the change in local atomic arrangements for $x>0.35$. This change in electronic structure exerts strong influence on physical properties, as demonstrated in the next Section.

**D. Physical properties**

Among the physical properties, the paramagnetic Pauli susceptibility of nonmagnetic alloys is directly related to the electronic structure. However, the magnetic susceptibility $\chi_{exp}$ of transition metals and alloys is quite complex and consists of three main contributions [38,49]:

$$\chi_{exp} = \chi_p + \chi_{dia} + \chi_{orb}, \qquad (6)$$

where $\chi_p$ is the Pauli paramagnetic contribution of d-electrons and $\chi_{dia}$ and $\chi_{orb}$ are diamagnetic and orbital paramagnetic contributions respectively. $\chi_{dia}$ and $\chi_{orb}$ are calculated by adding corresponding contributions from the constituents [49,83,84]. The Pauli paramagnetism of a d-band is enhanced over the free-electron value, $\chi_p^0 = \mu_0 \mu_B^2 N_0(E_F)$ (where $\mu_0$ is the permeability of the vacuum and $\mu_B$ is the Bohr magneton), by the exchange interaction, namely, $\chi_p = S\chi_p^0$ where $S$ is the Stoner enhancement factor, which also depends on $N_0(E_F)$. But in many amorphous TE-TL alloys [38] $S$ is nearly constant within their GFR, thus the variation of $\chi_p$ with concentration is dominated with $N_0(E_F)$. In spite of its complex structure $\chi_{exp}$ in a-TE-TL alloys often varies with composition in qualitatively the same way as $\gamma$ [38]. This is probably due to an approximately linear decrease of $\chi_{orb}$ with TL content, which does not affect the overall dependence on composition. This has also been observed in our alloys [36], therefore it seems to be quite general feature of non-magnetic a-TE-TL alloys which does not depend on their number of components. Indeed, when the variation of $\chi_{exp}$ with $x$ of our alloys is compared with those for binary amorphous alloys made of the same constituent TE and TL metals [22], the result is qualitatively the same as that for the corresponding values of $\gamma$ shown in Figure 5. In particular, the variation of $\chi_{exp}$ with $x$ for x ( Cu+Ni )≤0.51 is close to that for a-Zr-Ni alloys, but tends to saturate for $x>0.51$ (as does $\gamma$ in Figure 5). In Figure 6 we show that the variations of $\chi_p = \chi_{exp} - (\chi_{dia} + \chi_{orb})$ and $N_0(E_F)$ in our alloys are qualitatively the same, which shows that the change in electronic structure for $x \geq 0.35$ (*VEC*=7.4) affects the magnetic properties of our alloys.

As already noted, LTSH also provides information on the atomic vibrations and interatomic bonding [18,38,41]. This information is contained in the Debye $\beta T^3$ phonon term in $C_p$. Since $\beta \propto \theta_D^{-3}$ [41] one can calculate $\theta_D$ from the measured $\beta$. The variation of $\theta_D$ with $x$ in our alloys shown in Figure 7 is qualitatively the same as those of the thermal stability parameters $T_x$ and $T_l$ [36]: $\theta_D$ increases rapidly for $x \leq 0.35$, but tends to saturate in the Ni-rich region. Since the average atomic mass in the alloys studied changes linearly and relatively little with $x$ [18], the abrupt change in the variation of $\theta_D$ for $x>0.35$ is apparently related to a change in interatomic bonding and electronic structure, as was the case with $T_x$ and $T_l$ [36]. Careful measurements [18] have shown that $C_p$ increases faster with temperature than the $T^3$ law predicted by the Debye model [41]. This indicates the presence of the so-called "boson peak" (BP), an excess of low-energy vibrational states with respect to that predicted by the Debye model (e.g. [85]). A preliminary analysis [86] shows a non-monotonic variation of the magnitude of BP with Ni-content (inset to Figure 7) similar to that of $\theta_D$. To our knowledge this is the first observation of a possible correlation between the ordinary ($\theta_D$) and excess atomic vibrations in amorphous solids. At present we have no proper explanation for the observed behaviour of the BP in our alloys. However, a change in local atomic arrangements with Ni-concentration is expected to affect the atomic vibrations, thus the observed variations of both the BP and $\theta_D$ seem plausible. At present there is no commonly accepted explanation of the BP in glassy systems: some researchers emphasize localized phonon modes (like those caused by oscillations of loosely bonded atoms within the cages of surrounding atoms), whereas others ascribe the BP to a smeared van Hove anomaly [41, 85, 87]. We note however that the size of the BP depends quite strongly on the amount of quenched-in disorder which complicates the study of its variation with composition [86,87]. It is important to note that in our [18,36] and other a-TE-TL alloys (e.g. [38]) as well as in the crystalline cubic TE-based HEAs [21] the rule-of-mixtures does not describe $\theta_D$ properly. This conclusion holds even in the case of a self-consistent choice of the values of $\theta_D$ of the constituent elements [18].

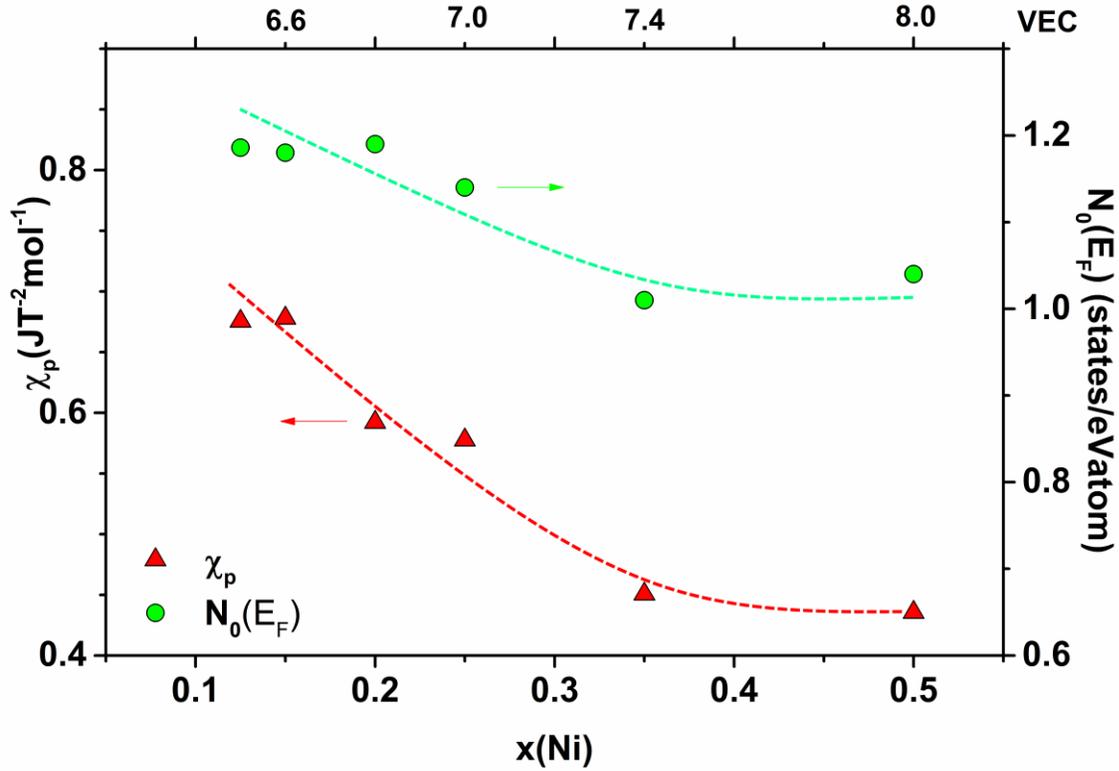

FIG. 6. Density of states $N_0(E_F)$ and Pauli susceptibility for $(TiZrNbCu)_{1-x}Ni_x$ alloys vs. $x$ and $VEC$ (upper abscissa). Left scale: $\chi_p$, right scale: $N_0(E_F)$. Dashed lines are guide for eye.

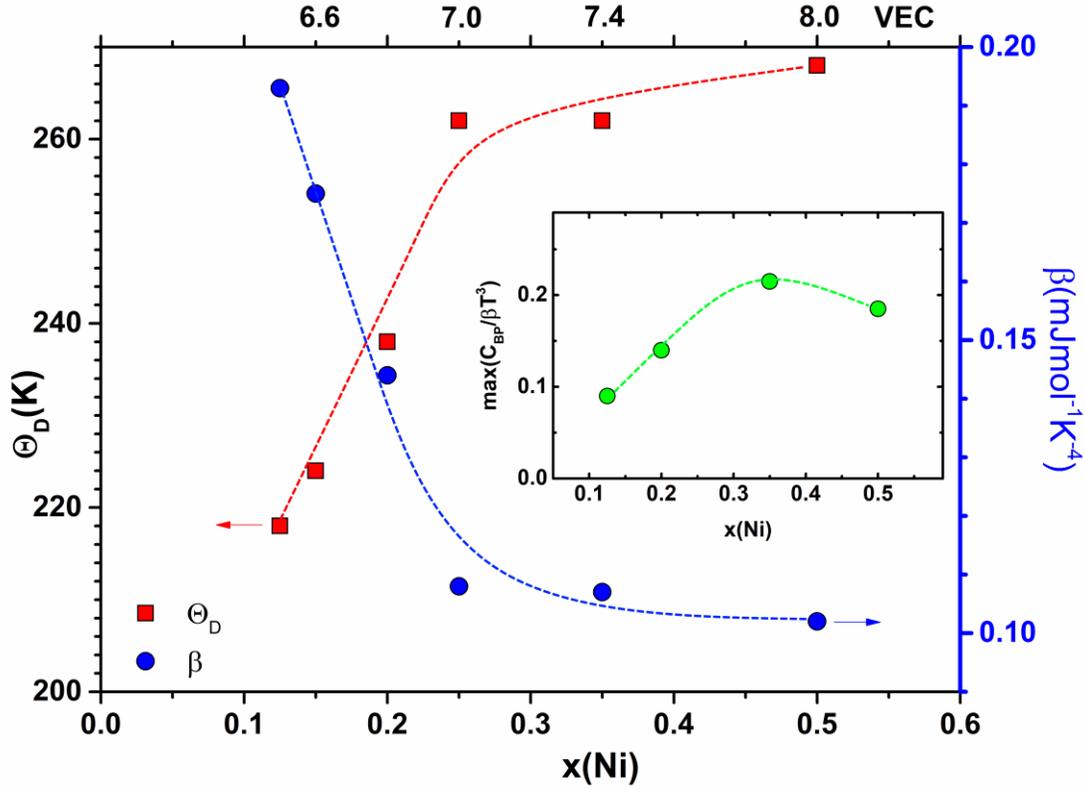

FIG. 7. Phonon coefficients $\beta$ and Debye temperatures of $(TiZrNbCu)_{1-x}Ni_x$ alloys vs. $x$ and *VEC* (upper abscissa). Left scale: variation of $\theta_D$ ; right scale: variation of $\beta$ . The inset: variation of the magnitude of boson peak [86] vs. $x$. Dashed lines are guide for eye.

As noted earlier [18, 36, 38, 47, 64, 85, 88] there is a very simple experimental relationship between electronic structure and mechanical properties (including hardness) and thermal stability (represented by $T_x$ and $T_l$) of the nonmagnetic a-TE-TL alloys. In particular, a decrease in $N_0(E_F)$ or $\gamma$ or $\chi_{exp}$ is usually accompanied by increases in Young`s modulus, hardness, Debye temperature and thermal stability of these alloys. Therefore, a decrease of $DoS$ at $E_F$ reflects in these systems an increase in the interatomic bonding, and accordingly the stiffness and parameters related to atomic vibrations and thermal stability increase too. We note that such a simple relationship between the electronic structure and interatomic bonding is quite common in crystalline alloys and in nonmagnetic a-TE-TL alloys it probably stems from their simple electronic band structure [64]. In Figure 8 we compare the variations of $E$ of our relaxed samples, that have received a short anneal close to $T_g$, with total (Cu+Ni)-content to those of binary a-TE-TL alloys composed from the same TEs and TLs [89-93]. Comparing the data in Figures 5, 6 and 8 we see that a correlation between electronic structure and $E$ is obeyed by our alloys, too. Further, the approximately linear variation of $E$ with $x$ in binary alloys seems to be replaced by a more complex variation in our multicomponent alloys. This probably reflects a change in the SRO and electronic structure for $x$(Cu+Ni)>0.51. The sensitivity of $E$ to quenched-in disorder and the degree of relaxation introduces considerable uncertainty in the variation of $E$ with $x$. As already noted [18, 36, 38] RoM provides a poor description of $E$ in all a-TE-TL alloys and it also fails to describe the mechanical properties of cubic crystalline TE-based HEAs [94].

For the sake of completeness, in what follows we briefly summarize some results of an ongoing study of the electronic transport properties of our alloys [81]. As is usual in a-TE-TL alloys [38], the electrical resistivities $\rho$ of a-(TiZrNbCu)$_{1-x}$Ni$_x$ alloys are high (>150 μΩcm) and accordingly decrease with increasing temperature over most of the explored temperature range (T≤300K). The variation of their electrical conductivities seems dominated by weak electron localization effects [95] over a broad temperature range (10 K-300 K) as is usual in a-TE-TL alloys [96, 97, 98]. All samples are superconducting, but with a low $T_c \leq 1.6K$. For $x \leq 0.35$ $T_c$ decreases approximately linearly with $x$, as is usual in nonmagnetic a-TE-TL alloys [99]. However, at higher $x=0.5$ $T_c$ tends to become constant, hence it follows the same trend with $x$ as $\gamma$ and $N_0(E_F)$, as is usual for disordered transition metal alloys [99]. A preliminary study of the Hall effect for $x \leq 0.25$ shows that the Hall coefficients $R_H$ are positive and change only a little with $x$, again in accord with the results for a-TE-TL alloys with lower TL contents [38,80]. Thus, the results for all physical properties of our multicomponent alloys studied until now seem similar to those of the corresponding binary alloys. This probably stems from their similar split-band electronic structures ( Figure 3).

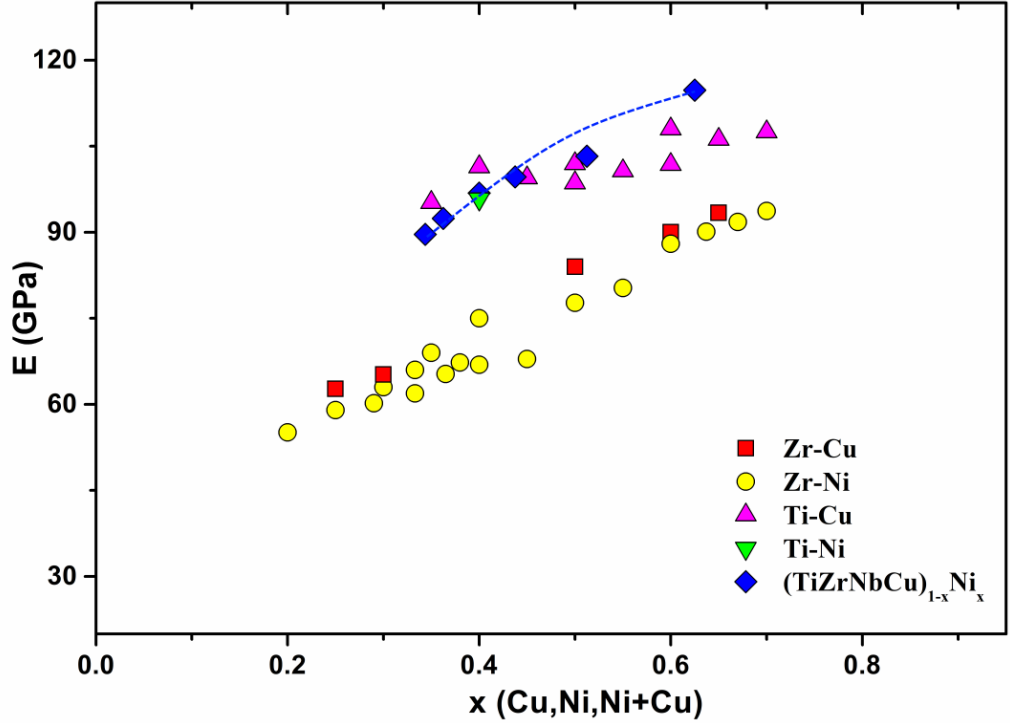

FIG. 8. Variation of $E$ vs. $x$ for (TiZrNbCu)$_{1-x}$ Ni$_x$ and some binary amorphous alloys. Data for Zr-Cu are from [89,90], for Zr-Ni from [89,91], for Ti-Cu from [88,92,93] and Ti-Ni from [90]. Dashed lines are guide for eye.

## IV.  CONCLUSION

The main results of comprehensive experimental studies of the relationship between the atomic structure, electronic structure and selected physical properties performed on well-characterized multicomponent (TiZrNbCu)$_{1-x}$Ni$_x$ ($x \leq 0.5$) amorphous alloys are presented. Owing to the broad concentration range explored, the relationship could be studied both in the high-entropy (a-HEA) region ($x \leq 0.25$) and in the region of conventional, Ni-based multicomponent amorphous alloys ($x \geq 0.35$). Such studies are important since the atomic structure and the corresponding electronic structure in metallic systems determine almost all their properties. Therefore, such studies provide a deeper understanding of the properties of the system investigated and moreover enables the prediction of some properties prior to measurement. We note that theoretical, ab-initio studies of the atomic structure-electronic structure-property relationship in multicomponent alloys are especially important and will greatly accelerate both the development and understanding of novel complex alloys. However, ab-initio studies of novel systems are complementary to experimental studies and cannot fully replace them in the development of novel materials.

The main result of our research is the complete consistency between studies of the atomic structure, electronic structure and selected physical properties. This clearly demonstrates the power of such an approach, because none of the correlations found can be properly described by using the rule-of-mixtures [18, 36].

As regards to the atomic structures, both the results of ordinary X-ray scattering and those from the ongoing study using synchrotron radiation provide evidence that the transition from the HEA composition range to that of conventional Ni-based amorphous alloys for $x \geq 0.35$ and an average number of valence electrons $VEC \geq 7.4$, is accompanied by a change in the interatomic arrangements. A possible cause of this change is the strong bonding tendency between Ni and Zr, Ti or Nb, as shown by the large negative values of $H_{mix}$ in Fig. 1. This results in the development of strong chemical short-range order which is reflected in the electronic structure [65]. The change results in an increase in the number of nearest neighbour atoms (obtained from the radial distribution function) and is therefore consistent with the change from the body centered cubic-to face centered cubic-like local atomic arrangements. A similar type of a change of crystal structure has been observed previously in crystalline HEAs for $VEC \geq 7$ [42,44].

Electronic structure, studied by the "low-temperature specific heat" and, for the first time, photoemission spectroscopy, also changes during the transition from the HEA to the conventional alloy region. In particular, the decrease of the electronic density of states at the Fermi level, $N_0(E_F)$ stops for $x \geq 0.35$. A possible cause is the change in the electronic states at the Fermi level from the dominant d-states of Zr and Ti to those of Ni, but further studies, including those of Hall effect and thermopower, are required in order to prove this conjecture.

As could be expected in metallic systems the selected physical properties reflect the electronic structure of our alloys. In particular the decrease of magnetic susceptibility, like that of $N_0(E_F)$, stops for $x \geq 0.35$. This also shows that the correlation between the electronic structure and selected physical properties in the amorphous alloys composed of early and late transition metals does not depend on the number of components of the alloy. The changes in the variations of the vibrational, elastic and thermal stability parameters with concentration for $x \geq 0.35$ show that a change in the interatomic arrangements is accompanied by a change in the interatomic bonding. In particular, the increase of all these parameters with $x$ stops, or slows down, for $x \geq 0.35$. This is consistent with the conjectured change from body-centered cubic to face-centered cubic-like local atomic arrangements.

Finally, we briefly address the relation between the present results for amorphous multi-component alloys of early and late transition metals with the corresponding results for binary amorphous alloys of Zr and Ti with Cu and Ni. As shown in sub-sections **C** and **D** due to the qualitatively similar split-band structure of the valence bands, both types of alloy show an approximately linear variation of a number of properties related to atomic and electronic structure, over a broad concentration range [36,38]. However, for binary systems with Cu, this linear variation is preserved over the entire glass forming range ($x \leq 0.9$), in binary alloys with Ni, the linear variation extends up to $x \geq 0.65$ ($VEC \geq 7.9$) while for our alloys it extends up to $VEC = 7.4$. The search for the origin of this difference is in progress.

**ACKNOWLEDGEMENT**

We thank Professor J. R. Cooper for useful suggestions and Professors E. Tafra and M. Basletić and Mr. M. Kuveždić for sharing their results prior to publication. Our research was supported by the project IZIP2016 of the University of Osijek. I. A. Figueroa acknowledges the financial support of UNAM-DGPA_PAPIIT, project No. IN101016. We thank the Diamond Light Source for access to beamline I12 that contributed to the results presented here. A. Lachova acknowledges financial support of the grant VEGA 1/0036/16.